\documentclass[a4paper,11pt]{article}
\pdfoutput=1 % if your are submitting a pdflatex (i.e. if you have
\usepackage{jcappub} % for details on the use of the package, please
\usepackage[T1]{fontenc} % if needed

\usepackage{float}

\usepackage{enumitem}

\newcommand{\Xmax}{X_{\rm max}}

\newcommand{\dEdX}{\mathrm{d}E/\mathrm{d}X}
\newcommand{\dEdXmax}{\left(\dEdX\right)_{\rm max}}
\newcommand{\gcm}{{\rm g/cm}^2}
\newcommand{\sigmaage}{\sigma_{\rm age}}

\title{Measurement of the average shape of longitudinal profiles of cosmic-ray air showers at the Pierre Auger Observatory}

\author{The Pierre Auger Collaboration}
\emailAdd{auger$\_$spokespersons@fnal.gov}

\abstract{The profile of the longitudinal development of showers produced by ultra-high energy cosmic rays carries information related to the interaction properties of the primary particles with atmospheric nuclei. In this work, we present the first measurement of the average shower profile in traversed atmospheric depth at the Pierre Auger Observatory. The shapes of profiles are well reproduced by the Gaisser-Hillas parametrization within the range studied, for $E>10^{17.8}$~eV. A detailed analysis of the systematic uncertainties is performed using 10 years of data and a full detector simulation. The average shape is quantified using two variables related to the width and asymmetry of the profile, and the results are compared with predictions of hadronic interaction models for different primary particles.}

\begin{document}
\maketitle
\flushbottom

\section{Introduction}
\label{sec:intro}

Ultra-High Energy Cosmic Rays (UHECRs) are the most energetic particles known in the Universe. The study of the cascades resulting from their interactions with atmospheric nuclei can provide a unique glimpse into hadronic interaction properties at center-of-mass energies more than one order of magnitude above those attained in human-made colliders. The collision of a UHECR with an atmospheric nucleus initiates an extensive air shower of secondary particles developing in the traversed air mass, usually referred to as slant depth, $X$. 

In a high-energy hadronic interaction, most of the secondary particles are pions, of which around one third are $\pi^{0}$ mesons. These immediately decay into two photons and initiate an electromagnetic cascade that is dominated by $e^{\pm}$ and $\gamma$. The charged pions, along with the other secondary hadrons produced in smaller numbers such as kaons and protons/neutrons, constitute the hadronic cascade. Since the interactions of $e^{\pm}$ and $\gamma$ yield virtually no hadrons, an air shower can be considered as the sum of two independent cascades: a hadronic cascade, waning as it penetrates further in the atmosphere losing energy via the decay of neutral pions, and an electromagnetic cascade that is constantly being fed by the hadronic counterpart. After only a few generations the vast majority of the total energy of the primary particle has been transferred to the electromagnetic part of the cascade. 

The cascade progresses until the average energy of single $e^{\pm}$ and $\gamma$ particles falls below the critical energy at which energy is lost predominantly by collisions instead of radiative processes. Atmospheric nitrogen molecules are excited by the passage of charged particles, and the subsequent nitrogen de-excitation results in the emission of a quantity of fluorescence light which is proportional to the energy lost by the shower electrons. The energy deposited by an air shower as a function of the traversed depth is known as its longitudinal profile, which can be measured by the detection of fluorescent light at the ground.

The integral of the longitudinal profile gives a calorimetric measurement of the shower energy. To measure this energy, we need to estimate the full profile. A functional form must be used to extrapolate outside the observed region. The atmospheric depth at which the profile has a maximum, $\Xmax$, is the variable most sensitive to the cross-section of the first interaction and to the mass of the primary particle. The precise shape of the energy deposit profile, however, has remained largely untested. This paper describes the first measurement of the shape of the longitudinal profile as a function of traversed atmospheric depth for UHECRs with energies above $10^{17.8}$~eV. While measured profiles for individual events often have large uncertainties, particularly at values far from the maximum, in this work a high precision is achieved by averaging, in energy bins, showers detected by the fluorescence detectors of the Pierre Auger Observatory.

The motivation for this measurement is three-fold. The first is to cross-check the assumption that shower profiles are well described by the currently used parametrization. The second is to provide a new way to control the quality of the shower reconstruction for the fluorescence detector. The third is to use the shape parameters to make new independent tests on hadronic interaction models and analysis of primary cosmic-ray composition.

This paper is organized as follows. The Pierre Auger Observatory and the event reconstruction procedure are described in section \ref{sec:event_rec}. In section \ref{sec:param}, the functional form used to fit the average profiles is defined, with its two variables and their interpretation. In section \ref{sec:data_sel}, we show how the average longitudinal profiles are reconstructed. The analysis is validated by comparing the average profiles obtained after full detector simulation and data reconstruction to the ones calculated directly from the simulated energy deposits in the atmosphere. The systematic uncertainties associated with the measurement are estimated in section \ref{sec:syst}. Finally, the results of the fit are presented in section \ref{sec:results}, and the shape variables measured at each energy are compared to the expectations for proton and iron initiated showers obtained from different hadronic interaction models. 

\section{Event reconstruction at the Pierre Auger Observatory}
\label{sec:event_rec}

The Pierre Auger Observatory \cite{bib:auger} is a hybrid detector, consisting of a 3000\,km$^2$ Surface Detector (SD) overlooked by the Fluorescence Detector (FD). The SD is composed of 1600 water-Cherenkov detectors separated by 1.5\,km. The FD consists of four sites with six telescopes each. The field of view of each telescope spans 30$^{\circ}$ in azimuth and ranges from 1.5$^{\circ}$ to 30$^{\circ}$ in elevation. Three additional telescopes called HEAT (High Elevation Auger Telescope) cover the elevation range from 30$^{\circ}$ to 60$^{\circ}$. This range is important for showers with energies lower than the ones studied in this paper and thus HEAT data were not used here. 

The measurement of atmospheric properties is essential for the reconstruction of the air showers measured by the detectors mentioned above. The molecular properties (temperature, humidity and pressure height profiles) are provided by the Global Data Assimilation System in three-hour intervals \cite{bib:GDAS}. The aerosol content is monitored hourly by calibrated laser shots from two laser facilities located near the center of the SD array, and cross-checked by LIDAR stations at each FD site. Cloud coverage in the shower path is measured by the LIDAR and cloud cameras at each site (every 15 and every 5 minutes, respectively), and complemented with data from the Geostationary Operational Environmental Satellite (GOES).
 
The reconstruction of shower profiles from FD data proceeds in the following steps (see e.g. \cite{bib:auger}). Firstly, the shower-detector plane, spanned by the pointing directions of pixels in the shower image, is calculated. The shower axis within this plane is obtained using the timing information of each pixel, as well as the timing of the closest SD station with signal (hybrid reconstruction).

For each time bin, a vector pointing from the telescope to the shower is defined, and the signals of all photomultipliers (PMTs) pointing to the same direction within a given opening angle are summed. This angle is determined event-by-event by maximizing the ratio of the signal to the accumulated noise from the night sky background. 

Given the reconstructed geometry, the signal at each time bin can be converted into an energy deposited by the shower at a given slant depth. Every time bin $i$ is projected to a path of length $l_i$ along the shower track. The slant depth, $X_i$, is inferred by integrating the atmospheric density through $l_i$. During its path from the shower axis to the FD, light is attenuated due to Rayleigh scattering on air and Mie scattering on aerosols. The light emitted on the shower track at time bin $i$ can be calculated from the measured light at the aperture corrected by this attenuation factor. 

The detected photons correspond to different light emission mechanisms and can reach the telescope directly or by scattering in the atmosphere. Fluorescence light is emitted isotropically along the shower track. High-energy charged particles emit Cherenkov light in a forward-concentrated beam. Even if the shower does not point directly to the detector, a fraction of this beam will be scattered into the field of view. This fraction is calculated taking into account the characteristics of both molecular and aerosol scattering in the atmosphere.

The Cherenkov and fluorescence light produced by an air shower are connected to the energy deposit by a set of linear equations \cite{bib:fit_met}. The profile of energy deposit as a function of slant depth is functionally described by the Gaisser-Hillas parametrization \cite{bib:gh},

\begin{equation}
  \label{usprl}
  f_\mathrm{GH}(X) = \dEdXmax\left( \frac{X-X_0}{\Xmax-X_0} \right)^{\frac{\Xmax-X_0}{\lambda}} \exp\left(\frac{\Xmax-X}{\lambda}\right),
\end{equation} 
which has four parameters: the maximum energy deposit, $\dEdXmax$, the depth at which this maximum is reached, $\Xmax$, and shape parameters $X_0$ and $\lambda$. This function is used in the calculation of the Cherenkov beam accumulated up to $X_i$, which determines the number of Cherenkov photons seen at the aperture. The proportionality between the number of fluorescence photons and the energy deposit is given by the fluorescence yield \cite{bib:flu_yield}, which depends on the molecular properties of the atmosphere. The statistical uncertainty in $\dEdX_i$ is calculated from the Poisson uncertainty of photoelectrons detected by the photomultipliers of the fluorescence telescopes.

The four parameters that describe the shower profile and their uncertainties are obtained from a log-likelihood fit to the number of photoelectrons detected at the PMTs. The calorimetric energy is the integral of $f_\mathrm{GH}$, and the total energy of the shower is estimated by correcting for the \emph{invisible} energy carried away by neutrinos and high-energy muons.

\begin{figure*}[tpb]
  \includegraphics[width=0.90\textwidth]{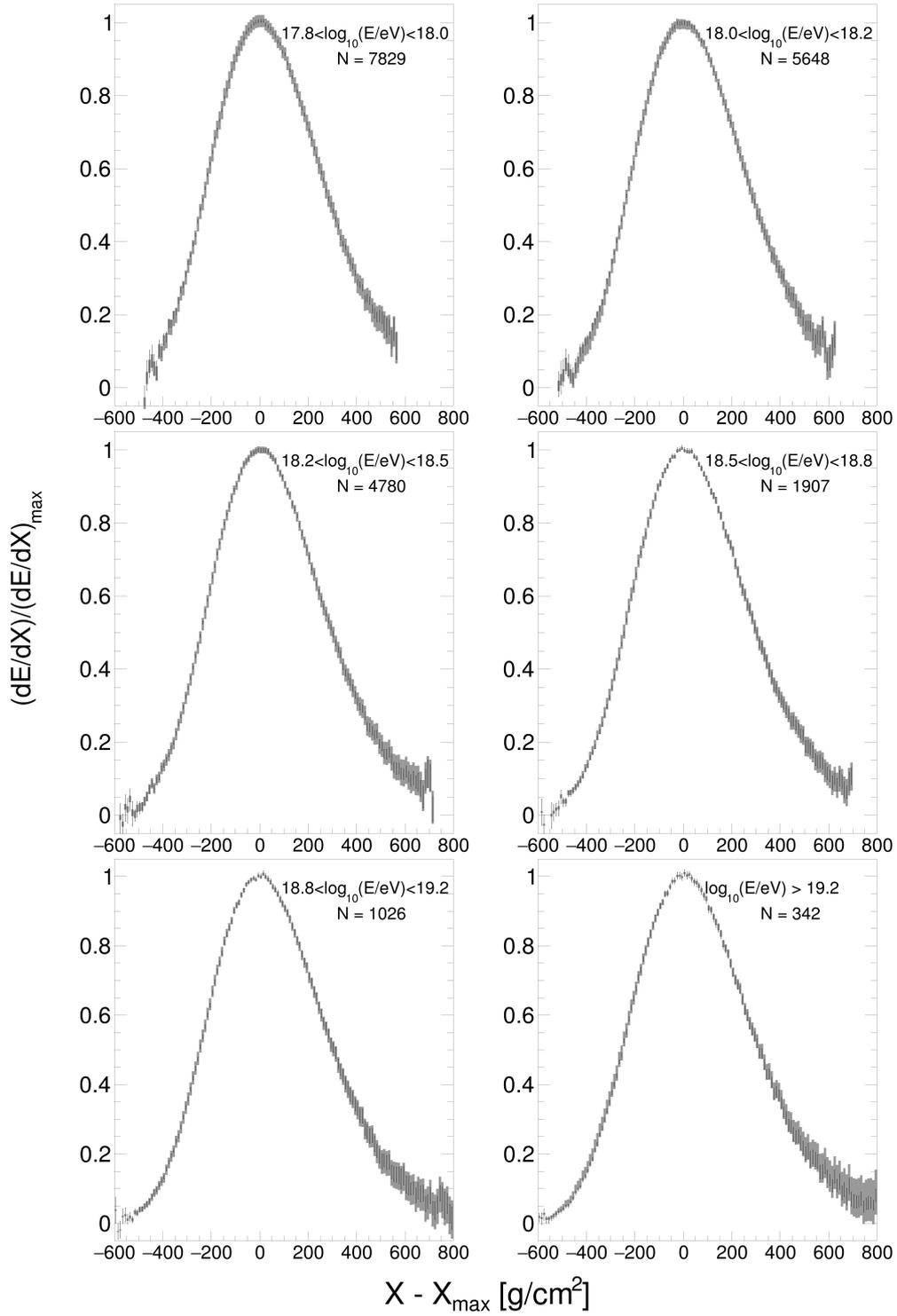} 
  \caption{Average longitudinal shower profiles for the selected data set, divided into the six energy bins used in this work (see legend for limits and number of events). The data is shown with statistical errors as a black line and an estimated bin-by-bin systematic uncertainty as a gray area.} 
  \label{fig:all_profiles}
\end{figure*}
\section{Average longitudinal shower profile}
\label{sec:param}

The maximum energy deposit of the longitudinal profile, $\dEdXmax$, is proportional to the energy of the primary particle and varies by three orders of magnitude in the energy range studied in this work. The $\Xmax$ value of each shower is, on average, a characteristic of the primary particle mass, but it varies greatly also for showers with the same primary, due to the depth at which the first interaction occurs, the large phase space for high-energy interactions and the general stochastic nature of shower development.

The focus of this work is to separate these two parameters, 
$\dEdXmax$ and $\Xmax$, to isolate the information contained in the profile shape itself. First, each measured shower profile is normalized to its fitted $\dEdXmax$. With this rescaling, all showers have a maximum value at 1. Then, the showers are transformed to the same development stage by translating the slant depth by $\Xmax$, i.e., $X'\equiv X - \Xmax$, thus all profiles are centered at zero. The normalized energy deposit profile, with $(\dEdX)' \equiv (\dEdX)/\dEdXmax$, can be described by the Gaisser-Hillas function, written as a function of parameters $R$ and $L$ \cite{bib:RL_first},

\begin{equation}
\label{usprl}
\left(\dEdX\right)' = \left(1+R \, \frac{X'}{L}\right)^{R^{-2}}
                \exp\left(-\frac{X'}{R \, L}\right),
\end{equation}
where $R=\sqrt{\lambda/|X'_0|}$, $L=\sqrt{|X'_0|\lambda}$ and $X'_0\equiv X_0 - \Xmax$. In this notation, the Gaisser-Hillas function is a Gaussian with standard deviation $L$, multiplied by a term that distorts it, with the asymmetry governed by $R$, and thus these parameters are less correlated than $\lambda$ and $X_0$. An equivalent parametrization as a function of the Full Width at Half Maximum, $f_{\rm FWHM}$, and asymmetry, $f$, has been reached independently \cite{bib:j_matthew}.

Note that in previous analyses by the HiRes/MIA \cite{bib:hires2001} and HiRes Collaborations \cite{bib:hires}, the energy deposit profile was constructed as a function of shower age ($s=3X/(X+2\Xmax)$), and the resulting shapes were found to be compatible with a Gaussian distribution with standard deviation $\sigmaage$. This width, however, is convolved with (and dominated by) the $\Xmax$ value\footnote{Shower age is defined as $s(X') = 1 + \frac{2X'}{3\Xmax - X'}$. As $L$ is the profile width in depth, a relation between $\sigmaage$ and $L$ can be found by making $X'=\pm L$ in the previous formula. Doing the derivative with respect to $L$ and $\Xmax$, we find that $\left( \frac{\partial \sigmaage}{\partial L} \Delta(L) \right) / \left(\frac{\partial \sigmaage}{\partial \Xmax} \Delta(\Xmax)\right) = \frac{\Xmax\Delta(L)}{L\Delta(\Xmax)} \approx 1/6$, where $\Delta$ stands for the proton-iron difference in each variable. So the majority of the composition separation in $\sigmaage$ comes from $\Xmax$.}. In this work we chose to represent the profiles in atmospheric depth because it preserves the measured event-by-event shape. Also, $R$ and $L$ taken from the fit to the profiles in atmospheric depth have been shown to be sensitive to the mass of the primary cosmic rays and to the properties of the high-energy interactions that occur at the start of the shower. The sensitivity is kept by an average profile \cite{bib:proc_ruben}, i.e., the average shape of a sample of profiles initiated by different primary particles is related to the average mass composition in the sample.

\section{Data selection and Monte Carlo validation}
\label{sec:data_sel}

The event selection used here is based on the selection criteria developed by the Pierre Auger Collaboration to measure the $\Xmax$ distributions \cite{bib:xmax}, and is applied to data covering the period from January 2004 to March 2015. Good weather conditions are required, with no clouds in the sky and a measured vertical aerosol optical depth lower than 0.1 at a height of 3~km.

A good shower profile reconstruction requires an accurate knowledge of the shower geometry, which is obtained by adding information from SD single-station triggers in coincidence with the FD shower trigger. An hybrid geometry reconstruction is performed using the start time of the pulse from the SD station with the largest signal in conjunction with the times from the triggered FD pixels. The SD station used should be within 1500 m of the core, and have an expected trigger probability for the specific geometry above 95\% for both proton and iron showers. For showers landing within the SD array, the closest station distance is usually within 750 m of the core, and the expected one-SD station trigger probability is 100\% for energies above $10^{17.6}$\,eV. So, the hybrid geometry requirement does not introduce any selection bias.

%A good hybrid geometry reconstruction is essential for a good shower profile reconstruction and it is obtained by using the start time of the pulse from the SD station with the largest signal in conjunction with the times from the triggered FD pixels. The SD station used should be within 1500~m of the core, and with an expected trigger probability for the specific geometry above 95\% for both proton and iron showers. For showers landing within the SD array, the closest station distance is usually within 750~m of the core, and the expected one-SD station trigger probability is 100\% for energies above 10$^{17.6}$~eV. So, such requirements do not introduce any selection bias.

On the measured longitudinal profile, strict cuts are made: at least 300\,$\gcm$ must be observed, including the $X_{\rm max}$ depth, for which the expected resolution must be less than 40\,$\gcm$. The minimum angle between the shower axis and the pointing vector of any of the pixels with signal has to be larger than 20$^{\circ}$, to minimize direct Cherenkov light that induces larger geometrical uncertainties in the determination of $\Xmax$. Finally, a fiducial field of view is defined to guarantee a uniform acceptance over a range in $\Xmax$ that covers the vast majority of the observed distributions in data, to reduce possible composition bias from specific geometries. 

In this work, two additional selection cuts with respect to the criteria developed in \cite{bib:xmax} were applied. When showers cross two telescopes, differences in alignment cause time residuals in the geometry fit, as the distance to the shower estimated by both telescopes is different. The estimation of the atmospheric attenuation depends on the distance and, therefore, also the energy deposit is affected. Large residuals were found in one of the telescopes (Coihueco 6), so events in which the profiles were measured using this telescope are excluded. Also, events for which the pixel time-fit used to determine the shower axis yielded very large reduced $\chi^{2}$ values (above 5) are excluded. The first cut affects approximately 3\% of the events (649), while in the second only two events are excluded.

In total, 21532 events are selected for analysis. These are divided into six energy bins. The shower profiles are constructed in 10\,$\gcm$ bins in $X'$, in which each energy deposit is accumulated with a weight corresponding to the inverse of its squared error. The average profiles for all energies are shown in figure \ref{fig:all_profiles}. 

\begin{figure}
\centering
\includegraphics[width=0.47\textwidth]{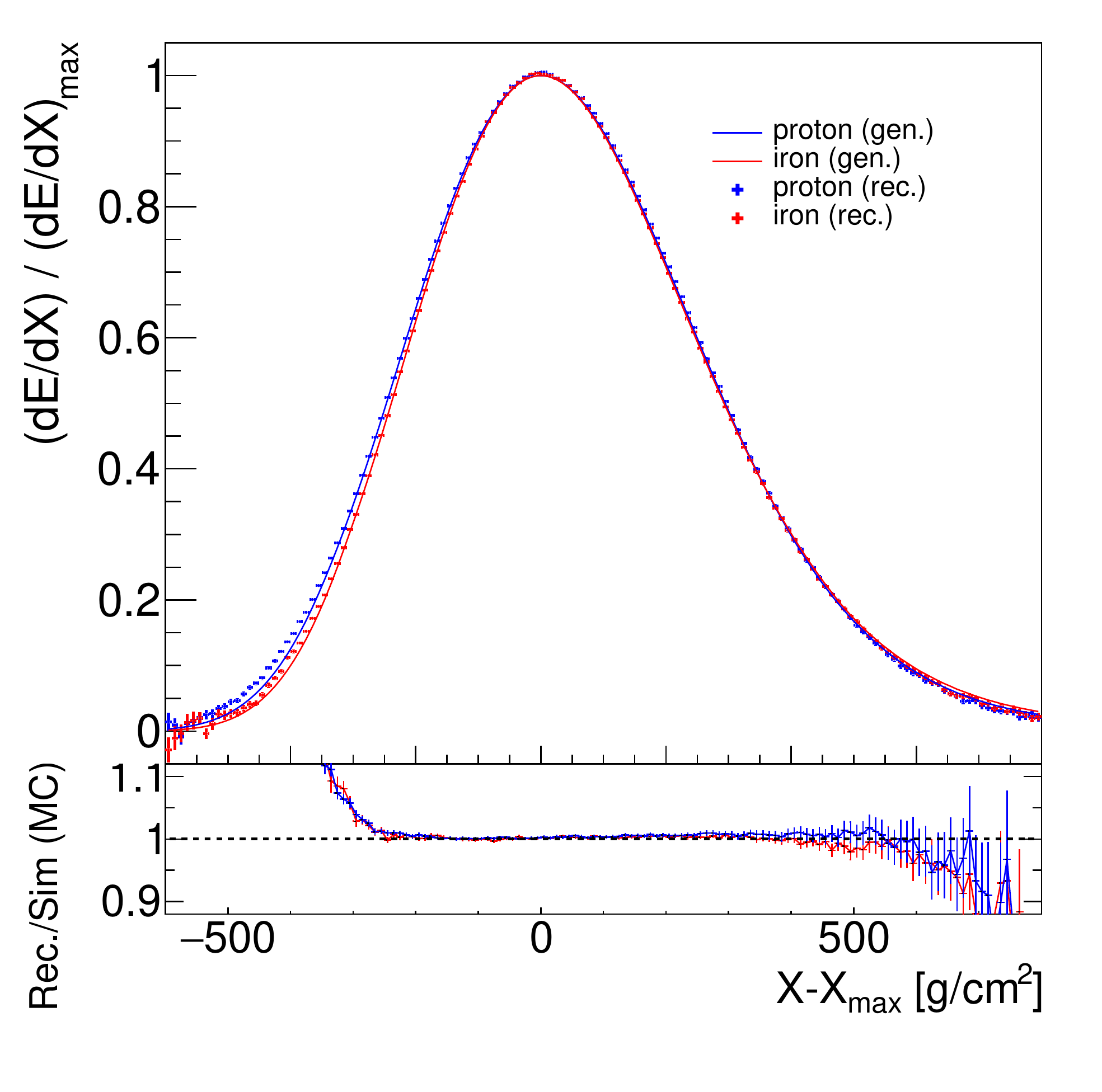}
\caption{\label{fig:aver_shower_mc} Average longitudinal shower profile for simulated events with energies between $10^{18.8}$ and $10^{19.2}$\,eV. The profiles for proton and iron are shown in blue and red, respectively. In the bottom plot, the ratio of reconstructed profiles to the generated ones is shown.}
\end{figure}

\begin{figure}
\centering
\includegraphics[width=0.47\textwidth]{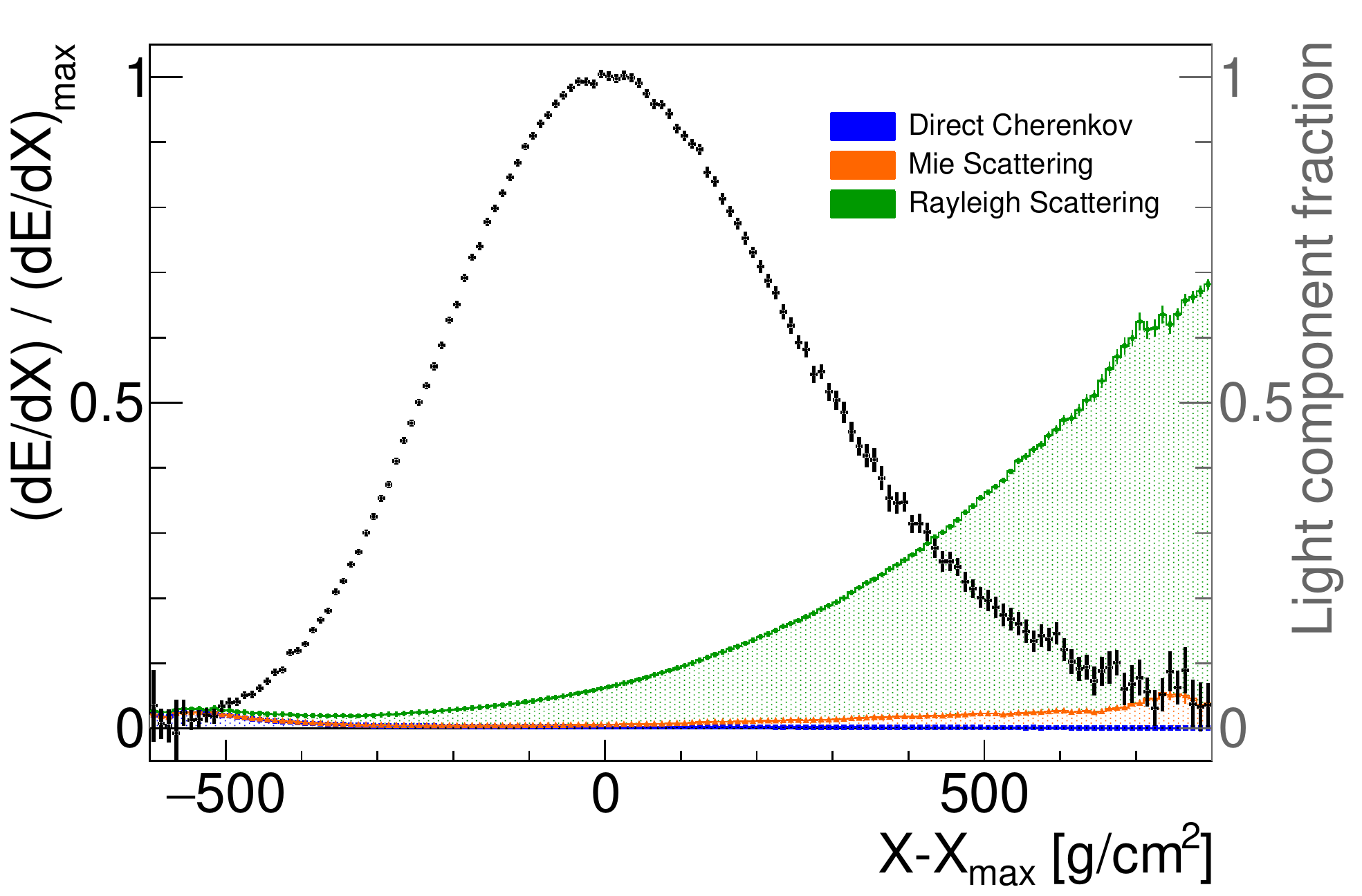}
\caption{\label{fig:aver_shower_data}Average longitudinal shower profile for data events with energies between $10^{18.8}$ and $10^{19.2}$\,eV. The measured normalized energy deposit is shown in black. The colored regions (detailed in the legend) represent the average fraction of direct and scattered Cherenkov light in the photons measured at the telescope aperture, computed in the individual shower reconstruction.}
\end{figure}

The reconstruction and analysis of the average longitudinal shower profiles was validated with a full detector simulation for energies between $10^{17.8}$ and $10^{20}$\,eV with proton and iron as primary particles. Comparing simulated and reconstructed showers an excellent agreement is found for the central part, but reconstruction increasingly deviates for negative $X'$, and a lower bound should be set for the analysis. At depths beyond $\Xmax$, showers lose most of the primary information, but the later parts of the profile are necessary to measure the profile width. An upper bound should maximize the relative importance of the profile start while keeping the statistical uncertainty on $L$ and $R$ below the proton/iron separation. While no detailed optimization was attempted for this first measurement, common bounds for all energies were set at [$-300$,$+200$]\,g/cm$^2$, fulfilling the above requirements. 

Figure \ref{fig:aver_shower_mc} shows the generated and reconstructed profiles for proton and iron simulations for energies around $10^{19}$\,eV; figure \ref{fig:aver_shower_data} shows the data reconstruction for the same energy bin, together with the different light components along the profile. Note that the chosen fit region is dominated by fluorescence light, directly proportional to the shower energy deposit along $X$.

The shower profiles are fitted with equation \ref{usprl}, leaving all parameters unconstrained. In addition to $R$ and $L$, the normalization and the maximum position are allowed to vary around $(\dEdX)'=1$ and $X'=0$\,$\gcm$ to account for possible effects of smearing due to $\Xmax$ resolution and corresponding bias in the average energy deposit, and for possible systematic effects in the extraction of $\Xmax$ and $\dEdXmax$ from the fit to single event profiles. The results for data are within the predictions from full-detector simulations of showers, i.e., at all energies, the normalization is unitary within 0.5\%, and the maximum position is within $\pm$1\,$\gcm$. 

To minimize the residual reconstruction biases, the differences between the reconstructed and generated profile parameters are calculated separately in proton and iron simulations, and the average across both primaries are subtracted from the values of $L$ and $R$ obtained in data. This bias correction is compatible with zero for energies above 1~EeV and is lower than 2\,$\gcm$ in $L$ and 0.005 in $R$ (with different signs for each primary), for the lowest energy bin analysed. Half of the applied correction is included as a systematic uncertainty.

\begin{figure}
\centering
\includegraphics[width=0.45\textwidth]{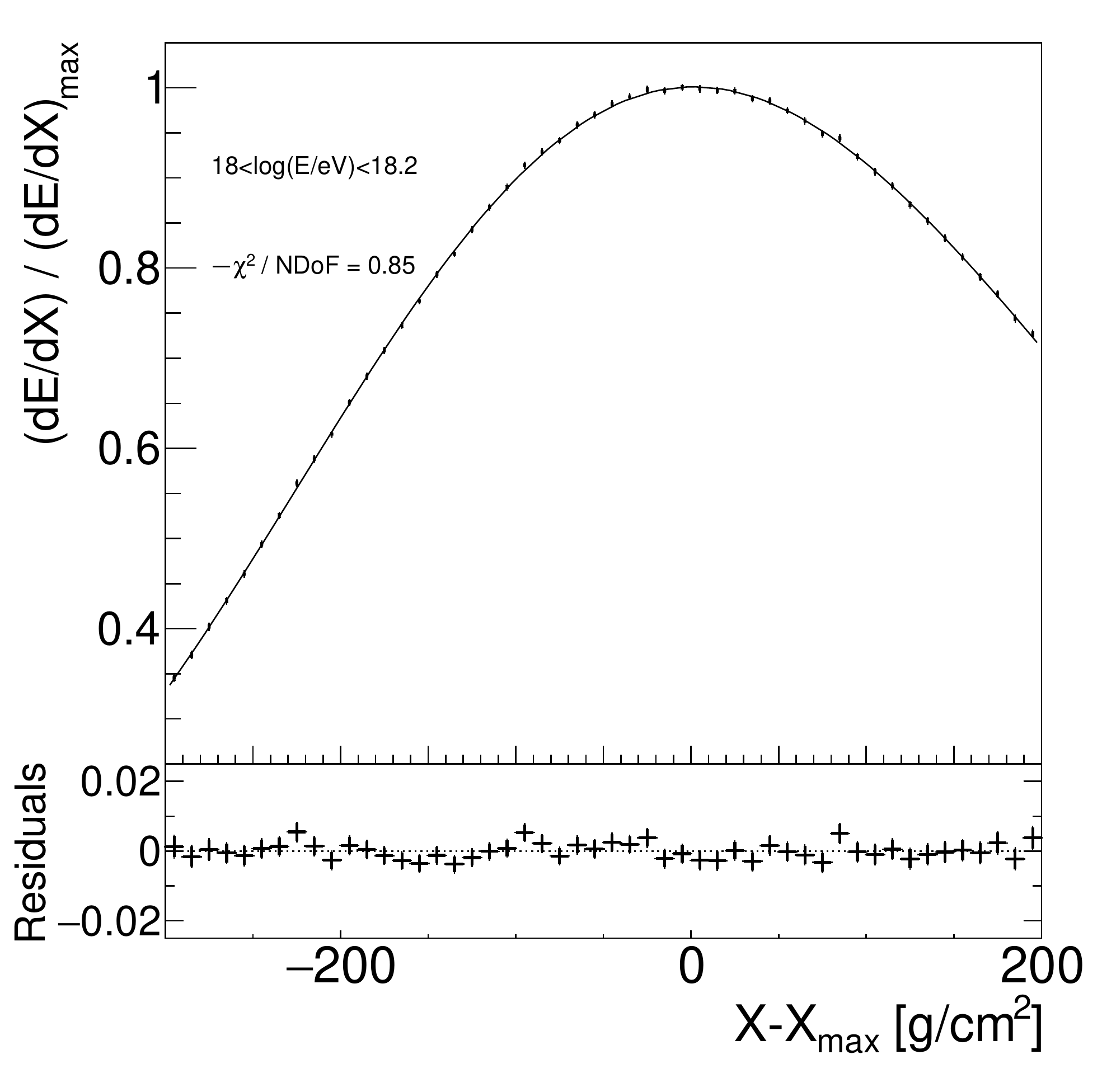}  
\includegraphics[width=0.45\textwidth]{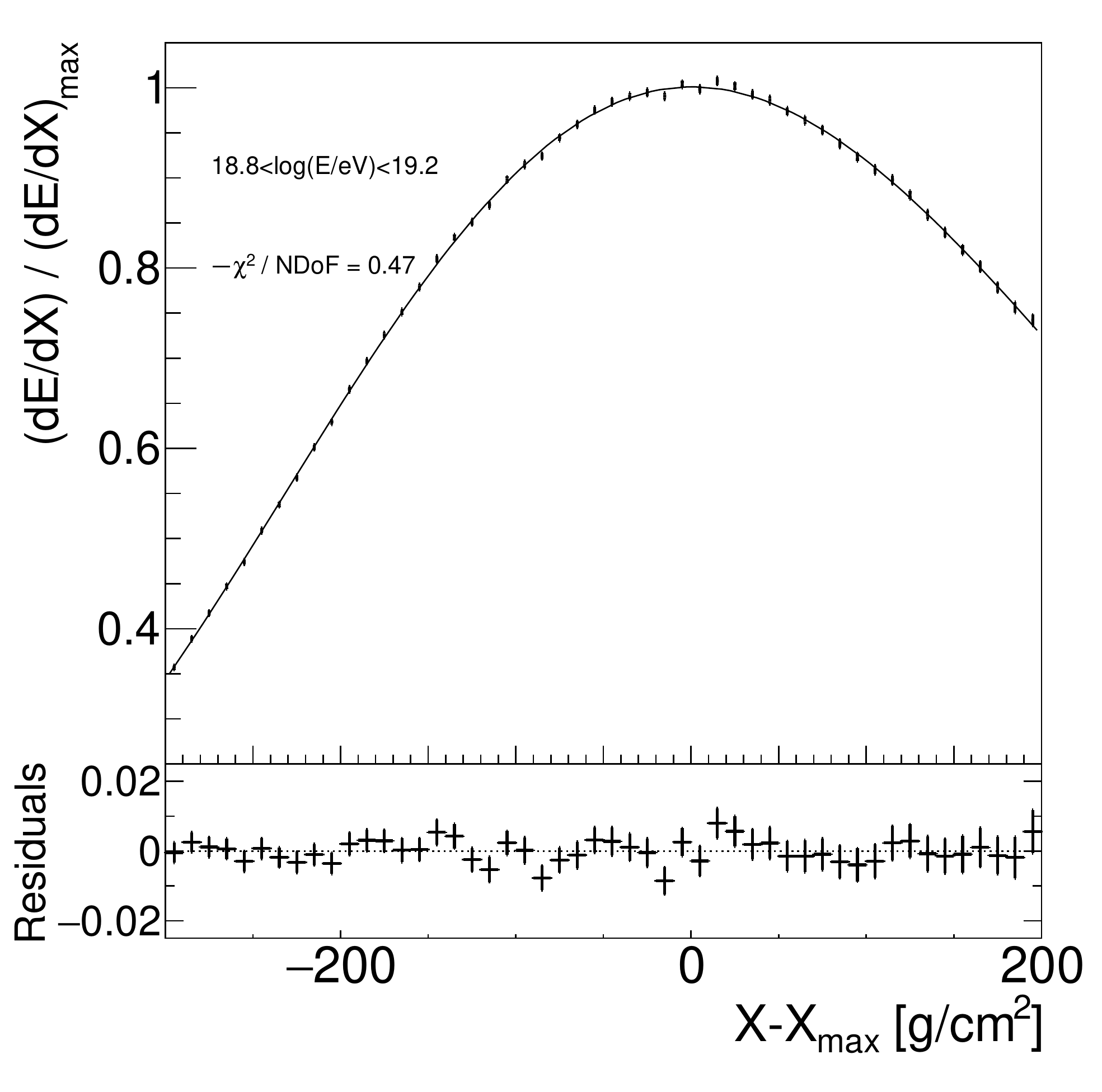}  
\caption{\label{fig:aver_shower} Measured average longitudinal shower profiles for energies between $10^{18}$ and $10^{18.2}$\,eV (left) and between $10^{18.8}$ and $10^{19.2}$\,eV (right). Data is shown in black, together with the Gaisser-Hillas fit to the profile. The residuals of the fit are shown in the bottom insets.}
\end{figure}

\section{Systematic uncertainties}
\label{sec:syst} 

The reconstruction of longitudinal profiles from the measurements at the FD requires several steps with which systematic uncertainties are associated. This section describes the effects we have considered, which are summarized in table \ref{table_syst}.

\begin{table}[tbp]
%{%\renewcommand{\arraystretch}{1.8}
\centering
\begin{tabular}{|l@{\hskip 0.3in}c@{\hskip 0.3in}c|}
 \hline 
%\hline%\toprule
                        & $\boldsymbol{R}$   & $\boldsymbol{L}$ [g/cm$^2$] \\
%\midrule
\hline
Atmosphere              & 0.030  & 5.5 \\ 
Light components \& fit & 0.018  & 3.3 \\ 
Geometry                & 0.018  & 2.2 \\
Detector                & 0.012  & 1.8 \\  
Bias correction \& Energy scale    & 0.010  & 1.0 \\ 
\textbf{Total}         & \textbf{0.040}  & \textbf{7.3} \\
\hline%\midrule
Statistical   & 0.012  & 0.9 \\
\hline%\bottomrule
\end{tabular}
%}
\caption{Breakdown of systematic uncertainties for $R$ and $L$. Uncertainties are energy dependent and asymmetric, so only the largest value is reported.}
\label{table_syst}
\end{table}

Atmospheric conditions play a crucial role in the propagation of the light from the shower to the fluorescence detector \cite{bib:atmos}. Several systematic uncertainties related to this process were studied. They included the impact of possible patches of clouds in the sky, differences found when separating data by the seasons of the year, and uncertainties in the overall aerosol content and its height dependence. The atmospheric aerosol attenuation, $\tau_A(h)$, is measured hourly with the central laser facility \cite{bib:clf,bib:aero}. The measurement compares the number of photons detected on the FD (as a function of height) in a given laser run with the one detected on a clear reference night. The aerosol height profile on the shower path can be calculated from $\tau_A(h)$, but has two main sources of uncertainty: the determination of the aerosol profile on the reference night (which is fixed for all showers) and the propagation of the uncertainty on the laser measurement at the FD to the path to each air shower (which varies from bin to bin and depends on geometry and atmospheric conditions). The uncertainty related to the estimation of the aerosol content was found to be the largest in this work, yielding approximately a $\pm0.02$ and $\pm5$\,$\gcm$ uncertainty in $R$ and $L$, respectively. These values are obtained by recalculating the average profiles after again reconstructing the full data sample, first with $\tau_A(h)$ changed coherently for all events according to the reference night uncertainties, and then with $\tau_A(h)$ varied within the non-correlated uncertainties of each event.

The uncertainties in the determination of the different light components were also considered. First, the reconstruction was repeated changing the fluorescence and Cherenkov yield values within their uncertainties (4\% and 5\%, respectively), and accounting or not for the multiple scattering corrections. Then, the data was separated according to the fluorescence fraction of the event, which gave a larger systematic difference. When only showers with fluorescence fractions lower than 90\% (the average value for the analyzed sample) are used, the resulting average profile is 2\,$\gcm$ larger in width, $L$, while the change in $R$ is negligible. 

The fit of the longitudinal profile of individual events is constrained in the integral of the normalized longitudinal profile, by an energy dependent value taken from shower simulations. Furthermore, the Gaisser-Hillas parameters are constrained by $X_0 = -121 \pm 172$\,$\gcm$ and $\lambda = 61 \pm 13$\,$\gcm$, values found previously in a small ensemble of very high-energy events with long tracks in the FD for which the unconstrained fit was possible \cite{bib:xmax}. The uncertainty in the estimation of the profile constraints was propagated in the reconstruction by shifting the central values by their standard deviation. This contribution is relatively more important in $L$ (around 3\,$\gcm$) than in $R$ (0.01).

The reconstructed shower geometry is given first by the zenith and azimuth angles of the shower-detector plane, and then the shower axis is defined by distance, angle and time references. All these five parameters are varied within $\pm1\sigma$ to obtain the systematic uncertainty associated with the reconstructed shower geometry. To ensure the quality of the reconstruction, the dependence of $R$ and $L$ on the zenith angle or distance from $X_{\rm max}$ to the telescope was also studied, but was found to be small and contained in the geometric systematic uncertainty of around 2\,$\gcm$ in $L$ and 0.02 in $R$.

The effect of the uncertainty in telescope alignment was also considered. One of the telescopes had been previously excluded from the analysis by looking directly at residuals in the timing at the crossing between telescopes. Smaller effects were tested by studying the telescope-to-telescope difference of the reconstructed shape. These are smaller than other systematic uncertainties, but at the level of 0.012 in $R$ and 1.8\,$\gcm$ in $L$.

Finally, the uncertainty of the energy scale of 14\% \cite{bib:verziAtICRC13} and the previously described proton-iron discrepancy from the bias correction are also added, but they are small (below 1\,$\gcm$ in $L$ and 0.01 in $R$) in comparison with the previously described ones. 

The uncertainties vary with energy and are asymmetric. Table \ref{table_syst} shows the largest value for each category, and table \ref{table.data.errors} shows the total error per energy bin.

\section{Results}
\label{sec:results}

\begin{figure*}[tbp]         
\centering
\includegraphics[width=0.47\textwidth]{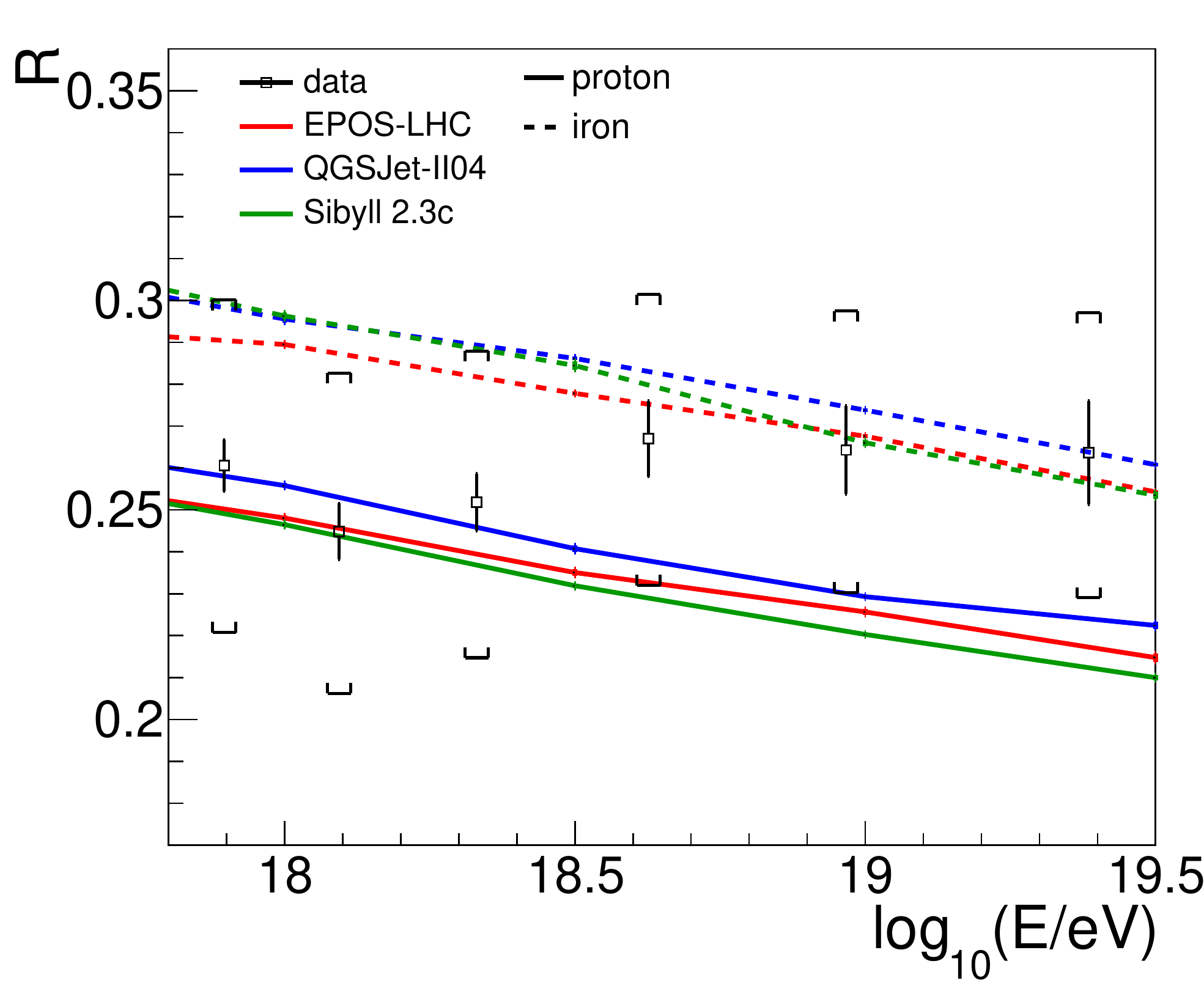}\hfill
\includegraphics[width=0.47\textwidth]{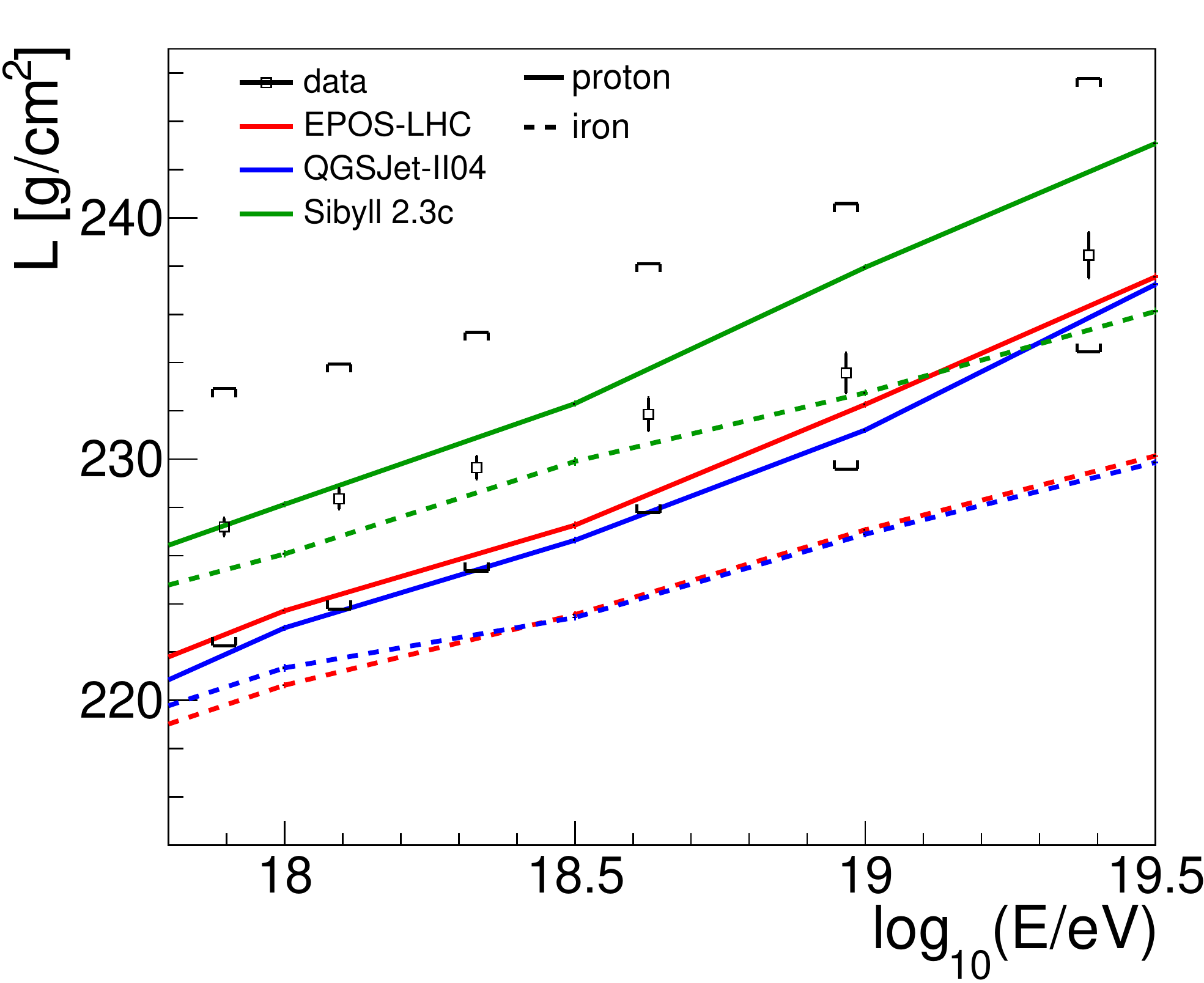}
\caption{$R$ (left) and $L$ (right) as a function of energy. The data are shown in black, with the vertical line representing the statistical error and the brackets the systematic uncertainty. Hadronic interaction models simulated with CORSIKA are shown (see legend), with full lines being proton and dashed lines iron predictions.}
\label{fig:canvas_rl}
\end{figure*}

\begin{table*}[tbp]
  \centering
  \renewcommand{\arraystretch}{1.5}
  %\begin{tabular*}{\textwidth}{|c@{\extracolsep{\fill}}@{\hskip 0.3in}|c@{\hskip 0.3in}|c@{\hskip 0.3in}|ccc@{\hskip 0.3in}|ccc|}
  \begin{tabular*}{\textwidth}{|c@{\extracolsep{\fill}}@{\hskip 0.1in}|c@{\hskip 0.1in}|c@{\hskip 0.1in}|ccc@{\hskip 0.1in}|ccc|}
\hline
%\multicolumn{ 1}{c}{} & \multicolumn{ 1}{c}{} & \multicolumn{ 1}{c}{} & \multicolumn{ 3}{c}{$R$}  &  \multicolumn{ 3}{c}{$L$ [$\gcm$]} \\
  &  &  & \multicolumn{ 3}{c|}{$\boldsymbol{R}$} &  \multicolumn{ 3}{c|}{$\boldsymbol{L}$ [$\gcm$]} \\ 
  Energy [eV]   &  $\left<\log_{10}[E/{\rm eV}]\right>$  & $N$ & $\left<value\right>$ & stat. & syst.     & $\left<value\right>$ & stat. & syst. \\ \hline
$10^{17.8}$ - $10^{18.0}$   &  17.90 & 7829 & 0.260 & 0.006 & $\substack{+0.039 \\ -0.040}$ & 226.2 & 0.4 & $\substack{+5.7 \\ -4.9}$ \\
$10^{18.0}$ - $10^{18.2}$   &  18.09 & 5648 & 0.244 & 0.007 & $\substack{+0.037 \\ -0.039}$ & 227.6 & 0.4 & $\substack{+5.6 \\ -4.5}$ \\
$10^{18.2}$ - $10^{18.5}$   &  18.33 & 4780 & 0.252 & 0.007 & $\substack{+0.035 \\ -0.037}$ & 229.1 & 0.5 & $\substack{+5.6 \\ -4.3}$ \\
$10^{18.5}$ - $10^{18.8}$   &  18.63 & 1907 & 0.267 & 0.009 & $\substack{+0.034 \\ -0.035}$ & 231.4 & 0.7 & $\substack{+6.2 \\ -4.1}$ \\
$10^{18.8}$ - $10^{19.2}$   &  18.97 & 1026 & 0.264 & 0.010 & $\substack{+0.033 \\ -0.034}$ & 233.3 & 0.8 & $\substack{+7.0 \\ -4.0}$ \\
$>$ $10^{19.2}$             &  19.38 & 342  & 0.264 & 0.012 & $\substack{+0.023 \\ -0.035}$ & 238.3 & 0.9 & $\substack{+7.3 \\ -4.0}$ \\
\hline
\end{tabular*} 
 \caption{\label{table.data.errors} $R$ and $L$ values for each of the measured energy bins, along with the statistical and systematic uncertainties.}
  \renewcommand{\arraystretch}{1}
\end{table*}

The fit of data profiles with the Gaisser-Hillas parametrization is shown in figure \ref{fig:aver_shower}. The fitted function follows the data points through the depth range used in this work, $[-300, +200]$~g/cm$^2$, with residuals within the statistical uncertainties. This is the first experimental demonstration that the Gaisser-Hillas parametrization is an accurate functional form to describe the central part of UHECR longitudinal profiles above $10^{17.8}$\,eV at 1\% accuracy. 

The values of $R$ and $L$ obtained from the fit for the six energy intervals studied are shown in table \ref{table.data.errors}, along with their energy dependent statistical and systematic uncertainties. The average energy and number of events, $N$, in each bin are also listed for reference.

The existing hadronic interaction models give different predictions for the shape variables. CORSIKA \cite{bib:corsika} was used to simulate proton, helium, nitrogen and iron showers with the EPOS-LHC \cite{bib:eposlhc}, QGSJetII-04 \cite{bib:qgsjetII04} and Sibyll2.3c \cite{bib:sibyll23} models. The evolution of $R$ and $L$ with energy, along with their respective systematic and statistical uncertainties, is shown in figure \ref{fig:canvas_rl}. Both the asymmetry, $R$, and the width, $L$, in data agree well with the predicted values for all models. For the asymmetry, all models give similar predictions, and the results seem to point to the composition becoming heavier with energy, although current systematic uncertainties still hinder any composition claim. For $L$, data is consistent with a linear increase with $\log_{10}(E/\mathrm{eV})$. $L$ is compatible with the predictions of Sibyll2.3c for all compositions, but points to a lighter composition if compared with the other two models, that predict smaller values of $L$.

\begin{figure*}         
\centering
\includegraphics[width=0.49\textwidth]{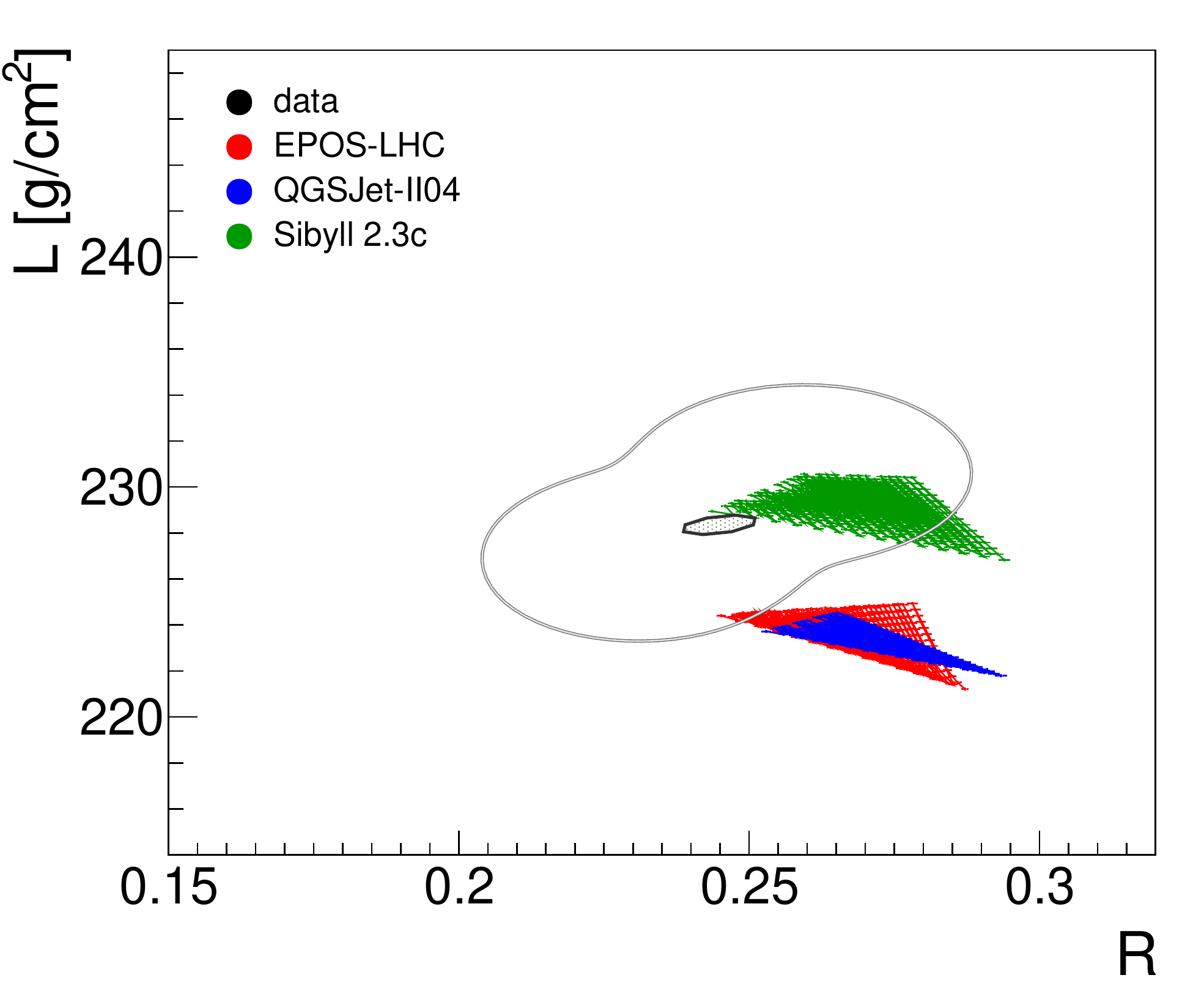}\hfil
\includegraphics[width=0.49\textwidth]{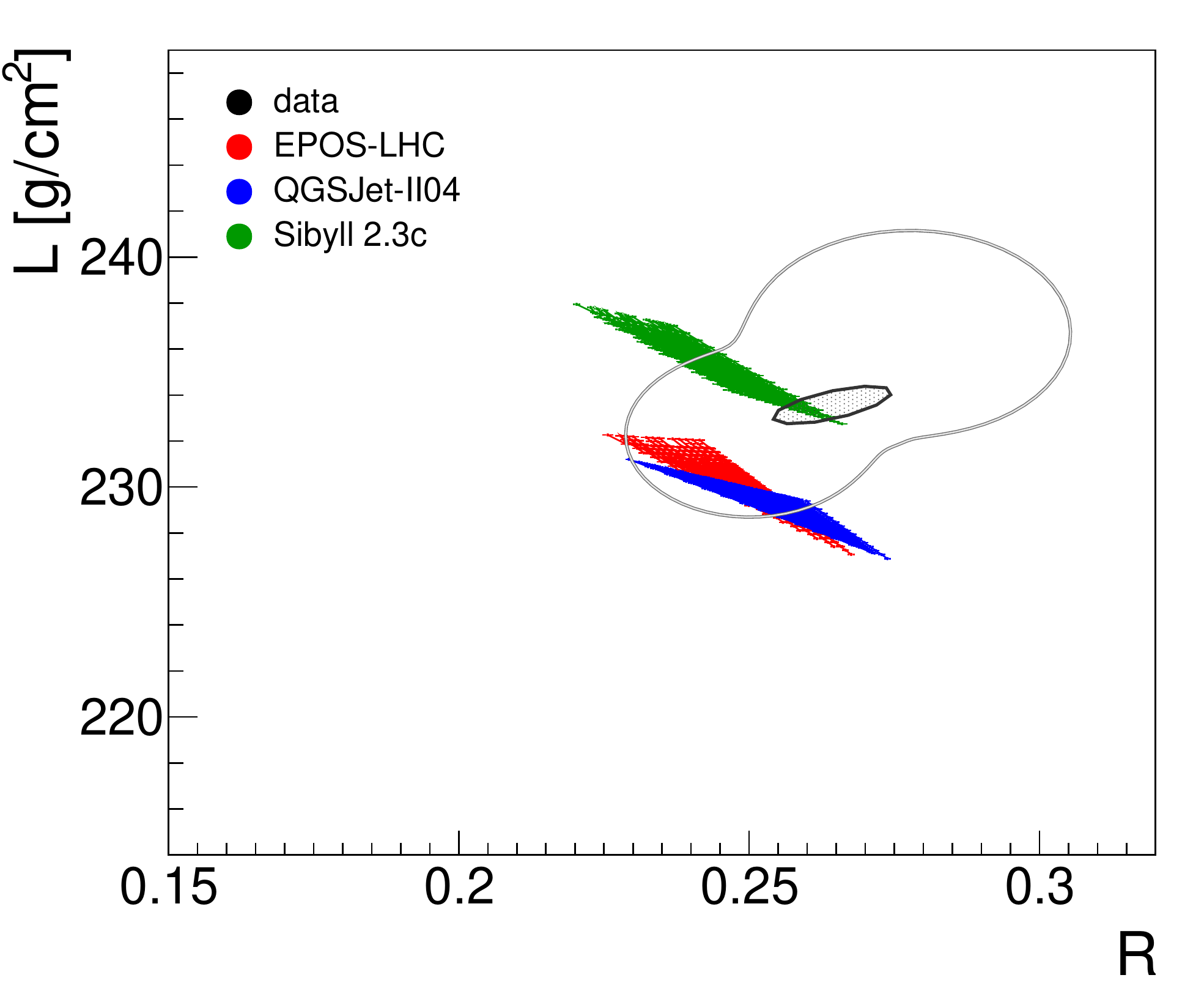}    
\caption{$L$ vs $R$ for the energy bin $10^{18}$ to $10^{18.2}$\,eV (left) and from $10^{18.8}$ to $10^{19.2}$\,eV (right). The inner dark grey ellipse shows the fitted value for data and its statistical uncertainty, and the outer light grey line the systematic uncertainty. For each hadronic model proton, helium, nitrogen and iron showers were simulated and average profiles were built making all possible combinations. Each of the points represents the value of $R$ and $L$ for a given model and composition combination, so the phase space spanned by each model is contained in its respective colored area. Pure proton is, for each model, on the upper left side (low $R$ and high $L$) and the transition to iron goes gradually to the lower right side.}
\label{fig:canvas_2d}
\end{figure*}

To understand better the interplay of the two measured variables, it is interesting to see the results in the ($R$, $L$) plane for two fixed energies (figure \ref{fig:canvas_2d}). In these plots, all composition scenarios are represented (as a combination of proton, He, N and Fe) for a given energy. For all models, proton has a lower $R$ and larger $L$ than iron, so moving from the top left to the bottom right points within a given model, makes the transition from lighter to heavier primaries. It is interesting, however, to note that the areas have different shapes: some models predict, at a given energy, a smaller $L$ for proton than for Helium, while others do the opposite.

Since $R$ and $L$ are experimentally correlated, these plots provide a smaller phase space within which to constrain the predictions of the hadronic interaction models. In figure \ref{fig:canvas_2d}, for the $10^{18}$\,eV energy bin (left), it can be seen that the average value in data is in the area occupied by most models for a light composition, while at $10^{19}$\,eV (right) it is within the predictions for intermediate mass primaries. 

Within the current experimental resolution, the data are fully compatible with most composition scenarios at the 2$\sigma$ level for all models. It is, however, interesting to note that Sibyll2.3c occupies a different phase space than the other two post-LHC models, indicating that a higher precision measurement of the profile shape could provide a test on the hadronic interaction models.

\section{Summary}

In this work, the Pierre Auger Observatory has been used to measure the average shape of the longitudinal profile of air showers. The method was first validated in a full detector simulation of proton and iron primaries, which showed that reconstructed and simulated profiles are in very good agreement for all energies above $10^{17.8}$ eV. The average longitudinal profiles as a function of atmospheric depth of ultra-high energy cosmic rays have been presented for the first time. They are well described by a Gaisser-Hillas function throughout the fitting range chosen, -300 to +200\,$\gcm$ around the shower maximum, to a 1\% level precision.
 
The systematic uncertainties contributing to the measurement were estimated, and we found that the uncertainty in the atmospheric aerosol content is the main factor affecting the shape of the reconstructed longitudinal profile.
In fact, the variations of $L$ and $R$ values obtained when dividing the sample according to different detection conditions are all within the range allowed by the atmospheric uncertainty. The validation of the profile shape around the shower maximum increases our confidence in its use for the determination of the shower calorimetric energy and of the $\Xmax$ parameter.

The two shape parameters measured in this work were compared with
model predictions and found to be compatible with most mass
composition scenarios at the two sigma level. However, different
hadronic models predict shape parameters which span different
regions in the ($R$, $L$) plane, even when considering all possible mass
compositions. Therefore, a future measurement of the average
longitudinal profile shape with smaller systematic uncertainties could
provide constraints on high-energy interaction models.

% created on 2018-11-09

\section*{Acknowledgments}

\begin{sloppypar}
The successful installation, commissioning, and operation of the Pierre
Auger Observatory would not have been possible without the strong
commitment and effort from the technical and administrative staff in
Malarg\"ue. We are very grateful to the following agencies and
organizations for financial support:
\end{sloppypar}

\begin{sloppypar}
Argentina -- Comisi\'on Nacional de Energ\'\i{}a At\'omica; Agencia Nacional de
Promoci\'on Cient\'\i{}fica y Tecnol\'ogica (ANPCyT); Consejo Nacional de
Investigaciones Cient\'\i{}ficas y T\'ecnicas (CONICET); Gobierno de la
Provincia de Mendoza; Municipalidad de Malarg\"ue; NDM Holdings and Valle
Las Le\~nas; in gratitude for their continuing cooperation over land
access; Australia -- the Australian Research Council; Brazil -- Conselho
Nacional de Desenvolvimento Cient\'\i{}fico e Tecnol\'ogico (CNPq);
Financiadora de Estudos e Projetos (FINEP); Funda\c{c}\~ao de Amparo \`a
Pesquisa do Estado de Rio de Janeiro (FAPERJ); S\~ao Paulo Research
Foundation (FAPESP) Grants No.~2010/07359-6 and No.~1999/05404-3;
Minist\'erio da Ci\^encia, Tecnologia, Inova\c{c}\~oes e Comunica\c{c}\~oes (MCTIC);
Czech Republic -- Grant No.~MSMT CR LTT18004, LO1305, LM2015038 and
CZ.02.1.01/0.0/0.0/16\_013/0001402; France -- Centre de Calcul
IN2P3/CNRS; Centre National de la Recherche Scientifique (CNRS); Conseil
R\'egional Ile-de-France; D\'epartement Physique Nucl\'eaire et Corpusculaire
(PNC-IN2P3/CNRS); D\'epartement Sciences de l'Univers (SDU-INSU/CNRS);
Institut Lagrange de Paris (ILP) Grant No.~LABEX ANR-10-LABX-63 within
the Investissements d'Avenir Programme Grant No.~ANR-11-IDEX-0004-02;
Germany -- Bundesministerium f\"ur Bildung und Forschung (BMBF); Deutsche
Forschungsgemeinschaft (DFG); Finanzministerium Baden-W\"urttemberg;
Helmholtz Alliance for Astroparticle Physics (HAP);
Helmholtz-Gemeinschaft Deutscher Forschungszentren (HGF); Ministerium
f\"ur Innovation, Wissenschaft und Forschung des Landes
Nordrhein-Westfalen; Ministerium f\"ur Wissenschaft, Forschung und Kunst
des Landes Baden-W\"urttemberg; Italy -- Istituto Nazionale di Fisica
Nucleare (INFN); Istituto Nazionale di Astrofisica (INAF); Ministero
dell'Istruzione, dell'Universit\'a e della Ricerca (MIUR); CETEMPS Center
of Excellence; Ministero degli Affari Esteri (MAE); M\'exico -- Consejo
Nacional de Ciencia y Tecnolog\'\i{}a (CONACYT) No.~167733; Universidad
Nacional Aut\'onoma de M\'exico (UNAM); PAPIIT DGAPA-UNAM; The Netherlands
-- Ministry of Education, Culture and Science; Netherlands Organisation
for Scientific Research (NWO); Dutch national e-infrastructure with the
support of SURF Cooperative; Poland -- National Centre for Research and
Development, Grant No.~ERA-NET-ASPERA/02/11; National Science Centre,
Grants No.~2013/08/M/ST9/00322, No.~2016/23/B/ST9/01635 and No.~HARMONIA
5--2013/10/M/ST9/00062, UMO-2016/22/M/ST9/00198; Portugal -- Portuguese
national funds and FEDER funds within Programa Operacional Factores de
Competitividade through Funda\c{c}\~ao para a Ci\^encia e a Tecnologia
(COMPETE); Romania -- Romanian Ministry of Research and Innovation
CNCS/CCCDI-UESFISCDI, projects
PN-III-P1-1.2-PCCDI-2017-0839/19PCCDI/2018, PN-III-P2-2.1-PED-2016-1922,
PN-III-P2-2.1-PED-2016-1659 and PN18090102 within PNCDI III; Slovenia --
Slovenian Research Agency; Spain -- Comunidad de Madrid; Fondo Europeo
de Desarrollo Regional (FEDER) funds; Ministerio de Econom\'\i{}a y
Competitividad; Xunta de Galicia; European Community 7th Framework
Program Grant No.~FP7-PEOPLE-2012-IEF-328826; USA -- Department of
Energy, Contracts No.~DE-AC02-07CH11359, No.~DE-FR02-04ER41300,
No.~DE-FG02-99ER41107 and No.~DE-SC0011689; National Science Foundation,
Grant No.~0450696; The Grainger Foundation; Marie Curie-IRSES/EPLANET;
European Particle Physics Latin American Network; European Union 7th
Framework Program, Grant No.~PIRSES-2009-GA-246806; and UNESCO.
\end{sloppypar}

\newpage

%\collaboration{The Pierre Auger Collaboration}
{\bf\Large{The Pierre Auger Collaboration}}\\
\\
% created on 2018-11-09
A.~Aab$^{75}$,
P.~Abreu$^{67}$,
M.~Aglietta$^{50,49}$,
I.F.M.~Albuquerque$^{19}$,
J.M.~Albury$^{12}$,
I.~Allekotte$^{1}$,
A.~Almela$^{8,11}$,
J.~Alvarez Castillo$^{63}$,
J.~Alvarez-Mu\~niz$^{74}$,
G.A.~Anastasi$^{42,43}$
% ...

% created on 2018-11-09

% needs \usepackage{enumitem}
\begin{description}[labelsep=0.2em,align=right,labelwidth=0.7em,labelindent=0em,leftmargin=2em,noitemsep]
\item[$^{1}$] Centro At\'omico Bariloche and Instituto Balseiro (CNEA-UNCuyo-CONICET), San Carlos de Bariloche, Argentina
\item[$^{2}$] Centro de Investigaciones en L\'aseres y Aplicaciones, CITEDEF and CONICET, Villa Martelli, Argentina
\item[$^{3}$] Departamento de F\'\i{}sica and Departamento de Ciencias de la Atm\'osfera y los Oc\'eanos, FCEyN, Universidad de Buenos Aires and CONICET, Buenos Aires, Argentina
\item[$^{4}$] IFLP, Universidad Nacional de La Plata and CONICET, La Plata, Argentina
\item[$^{5}$] Instituto de Astronom\'\i{}a y F\'\i{}sica del Espacio (IAFE, CONICET-UBA), Buenos Aires, Argentina
\item[$^{6}$] Instituto de F\'\i{}sica de Rosario (IFIR) -- CONICET/U.N.R.\ and Facultad de Ciencias Bioqu\'\i{}micas y Farmac\'euticas U.N.R., Rosario, Argentina
\item[$^{7}$] Instituto de Tecnolog\'\i{}as en Detecci\'on y Astropart\'\i{}culas (CNEA, CONICET, UNSAM), and Universidad Tecnol\'ogica Nacional -- Facultad Regional Mendoza (CONICET/CNEA), Mendoza, Argentina
\item[$^{8}$] Instituto de Tecnolog\'\i{}as en Detecci\'on y Astropart\'\i{}culas (CNEA, CONICET, UNSAM), Buenos Aires, Argentina
\item[$^{9}$] Observatorio Pierre Auger, Malarg\"ue, Argentina
\item[$^{10}$] Observatorio Pierre Auger and Comisi\'on Nacional de Energ\'\i{}a At\'omica, Malarg\"ue, Argentina
\item[$^{11}$] Universidad Tecnol\'ogica Nacional -- Facultad Regional Buenos Aires, Buenos Aires, Argentina
\item[$^{12}$] University of Adelaide, Adelaide, S.A., Australia
\item[$^{13}$] Universit\'e Libre de Bruxelles (ULB), Brussels, Belgium
\item[$^{14}$] Vrije Universiteit Brussels, Brussels, Belgium
\item[$^{15}$] Centro Brasileiro de Pesquisas Fisicas, Rio de Janeiro, RJ, Brazil
\item[$^{16}$] Centro Federal de Educa\c{c}\~ao Tecnol\'ogica Celso Suckow da Fonseca, Nova Friburgo, Brazil
\item[$^{17}$] Universidade de S\~ao Paulo, Escola de Engenharia de Lorena, Lorena, SP, Brazil
\item[$^{18}$] Universidade de S\~ao Paulo, Instituto de F\'\i{}sica de S\~ao Carlos, S\~ao Carlos, SP, Brazil
\item[$^{19}$] Universidade de S\~ao Paulo, Instituto de F\'\i{}sica, S\~ao Paulo, SP, Brazil
\item[$^{20}$] Universidade Estadual de Campinas, IFGW, Campinas, SP, Brazil
\item[$^{21}$] Universidade Estadual de Feira de Santana, Feira de Santana, Brazil
\item[$^{22}$] Universidade Federal do ABC, Santo Andr\'e, SP, Brazil
\item[$^{23}$] Universidade Federal do Paran\'a, Setor Palotina, Palotina, Brazil
\item[$^{24}$] Universidade Federal do Rio de Janeiro, Instituto de F\'\i{}sica, Rio de Janeiro, RJ, Brazil
\item[$^{25}$] Universidade Federal do Rio de Janeiro (UFRJ), Observat\'orio do Valongo, Rio de Janeiro, RJ, Brazil
\item[$^{26}$] Universidade Federal Fluminense, EEIMVR, Volta Redonda, RJ, Brazil
\item[$^{27}$] Universidad de Medell\'\i{}n, Medell\'\i{}n, Colombia
\item[$^{28}$] Universidad Industrial de Santander, Bucaramanga, Colombia
\item[$^{29}$] Charles University, Faculty of Mathematics and Physics, Institute of Particle and Nuclear Physics, Prague, Czech Republic
\item[$^{30}$] Institute of Physics of the Czech Academy of Sciences, Prague, Czech Republic
\item[$^{31}$] Palacky University, RCPTM, Olomouc, Czech Republic
\item[$^{32}$] Institut de Physique Nucl\'eaire d'Orsay (IPNO), Universit\'e Paris-Sud, Univ.\ Paris/Saclay, CNRS-IN2P3, Orsay, France
\item[$^{33}$] Laboratoire de Physique Nucl\'eaire et de Hautes Energies (LPNHE), Universit\'es Paris 6 et Paris 7, CNRS-IN2P3, Paris, France
\item[$^{34}$] Univ.\ Grenoble Alpes, CNRS, Grenoble Institute of Engineering Univ.\ Grenoble Alpes, LPSC-IN2P3, 38000 Grenoble, France, France
\item[$^{35}$] Bergische Universit\"at Wuppertal, Department of Physics, Wuppertal, Germany
\item[$^{36}$] Karlsruhe Institute of Technology, Institute for Experimental Particle Physics (ETP), Karlsruhe, Germany
\item[$^{37}$] Karlsruhe Institute of Technology, Institut f\"ur Kernphysik, Karlsruhe, Germany
\item[$^{38}$] Karlsruhe Institute of Technology, Institut f\"ur Prozessdatenverarbeitung und Elektronik, Karlsruhe, Germany
\item[$^{39}$] RWTH Aachen University, III.\ Physikalisches Institut A, Aachen, Germany
\item[$^{40}$] Universit\"at Hamburg, II.\ Institut f\"ur Theoretische Physik, Hamburg, Germany
\item[$^{41}$] Universit\"at Siegen, Fachbereich 7 Physik -- Experimentelle Teilchenphysik, Siegen, Germany
\item[$^{42}$] Gran Sasso Science Institute, L'Aquila, Italy
\item[$^{43}$] INFN Laboratori Nazionali del Gran Sasso, Assergi (L'Aquila), Italy
\item[$^{44}$] INFN, Sezione di Catania, Catania, Italy
\item[$^{45}$] INFN, Sezione di Lecce, Lecce, Italy
\item[$^{46}$] INFN, Sezione di Milano, Milano, Italy
\item[$^{47}$] INFN, Sezione di Napoli, Napoli, Italy
\item[$^{48}$] INFN, Sezione di Roma ``Tor Vergata'', Roma, Italy
\item[$^{49}$] INFN, Sezione di Torino, Torino, Italy
\item[$^{50}$] Osservatorio Astrofisico di Torino (INAF), Torino, Italy
\item[$^{51}$] Politecnico di Milano, Dipartimento di Scienze e Tecnologie Aerospaziali , Milano, Italy
\item[$^{52}$] Universit\`a del Salento, Dipartimento di Matematica e Fisica ``E.\ De Giorgi'', Lecce, Italy
\item[$^{53}$] Universit\`a dell'Aquila, Dipartimento di Scienze Fisiche e Chimiche, L'Aquila, Italy
\item[$^{54}$] Universit\`a di Catania, Dipartimento di Fisica e Astronomia, Catania, Italy
\item[$^{55}$] Universit\`a di Milano, Dipartimento di Fisica, Milano, Italy
\item[$^{56}$] Universit\`a di Napoli ``Federico II'', Dipartimento di Fisica ``Ettore Pancini'', Napoli, Italy
\item[$^{57}$] Universit\`a di Roma ``Tor Vergata'', Dipartimento di Fisica, Roma, Italy
\item[$^{58}$] Universit\`a Torino, Dipartimento di Fisica, Torino, Italy
\item[$^{59}$] Benem\'erita Universidad Aut\'onoma de Puebla, Puebla, M\'exico
\item[$^{60}$] Centro de Investigaci\'on y de Estudios Avanzados del IPN (CINVESTAV), M\'exico, D.F., M\'exico
\item[$^{61}$] Unidad Profesional Interdisciplinaria en Ingenier\'\i{}a y Tecnolog\'\i{}as Avanzadas del Instituto Polit\'ecnico Nacional (UPIITA-IPN), M\'exico, D.F., M\'exico
\item[$^{62}$] Universidad Aut\'onoma de Chiapas, Tuxtla Guti\'errez, Chiapas, M\'exico
\item[$^{63}$] Universidad Nacional Aut\'onoma de M\'exico, M\'exico, D.F., M\'exico
\item[$^{64}$] Institute of Nuclear Physics PAN, Krakow, Poland
\item[$^{65}$] University of \L{}\'od\'z, Faculty of Astrophysics, \L{}\'od\'z, Poland
\item[$^{66}$] University of \L{}\'od\'z, Faculty of High-Energy Astrophysics,\L{}\'od\'z, Poland
\item[$^{67}$] Laborat\'orio de Instrumenta\c{c}\~ao e F\'\i{}sica Experimental de Part\'\i{}culas -- LIP and Instituto Superior T\'ecnico -- IST, Universidade de Lisboa -- UL, Lisboa, Portugal
\item[$^{68}$] ``Horia Hulubei'' National Institute for Physics and Nuclear Engineering, Bucharest-Magurele, Romania
\item[$^{69}$] Institute of Space Science, Bucharest-Magurele, Romania
\item[$^{70}$] University Politehnica of Bucharest, Bucharest, Romania
\item[$^{71}$] Center for Astrophysics and Cosmology (CAC), University of Nova Gorica, Nova Gorica, Slovenia
\item[$^{72}$] Experimental Particle Physics Department, J.\ Stefan Institute, Ljubljana, Slovenia
\item[$^{73}$] Universidad de Granada and C.A.F.P.E., Granada, Spain
\item[$^{74}$] Instituto Galego de F\'\i{}sica de Altas Enerx\'\i{}as (I.G.F.A.E.), Universidad de Santiago de Compostela, Santiago de Compostela, Spain
\item[$^{75}$] IMAPP, Radboud University Nijmegen, Nijmegen, The Netherlands
\item[$^{76}$] KVI -- Center for Advanced Radiation Technology, University of Groningen, Groningen, The Netherlands
\item[$^{77}$] Nationaal Instituut voor Kernfysica en Hoge Energie Fysica (NIKHEF), Science Park, Amsterdam, The Netherlands
\item[$^{78}$] Stichting Astronomisch Onderzoek in Nederland (ASTRON), Dwingeloo, The Netherlands
\item[$^{79}$] Universiteit van Amsterdam, Faculty of Science, Amsterdam, The Netherlands
\item[$^{80}$] Case Western Reserve University, Cleveland, OH, USA
\item[$^{81}$] Colorado School of Mines, Golden, CO, USA
\item[$^{82}$] Department of Physics and Astronomy, Lehman College, City University of New York, Bronx, NY, USA
\item[$^{83}$] Louisiana State University, Baton Rouge, LA, USA
\item[$^{84}$] Michigan Technological University, Houghton, MI, USA
\item[$^{85}$] New York University, New York, NY, USA
\item[$^{86}$] Pennsylvania State University, University Park, PA, USA
\item[$^{87}$] University of Chicago, Enrico Fermi Institute, Chicago, IL, USA
\item[$^{88}$] University of Delaware, Department of Physics and Astronomy, Newark, USA
\item[$^{89}$] University of Nebraska, Lincoln, NE, USA
\item[] -----
\item[$^{a}$] School of Physics and Astronomy, University of Leeds, Leeds, United Kingdom
\item[$^{b}$] Max-Planck-Institut f\"ur Radioastronomie, Bonn, Germany
\item[$^{c}$] Fermi National Accelerator Laboratory, USA
\item[$^{d}$] also at Universidade Federal de Alfenas, Po\c{c}os de Caldas, Brazil
\item[$^{e}$] Colorado State University, Fort Collins, CO, USA
\item[$^{f}$] now at Institute for Cosmic Ray Research, University of Tokyo
\item[$^{g}$] also at Karlsruhe Institute of Technology, Karlsruhe, Germany
\item[$^{h}$] also at University of Bucharest, Physics Department, Bucharest, Romania
\end{description}


\begin{thebibliography}{99}

  \def\arxiv#1{[\href{http://arxiv.org/abs/#1}{arXiv:#1}]}

\bibitem{bib:auger}
Pierre Auger Collaboration, A.~Aab \textit{et al.}, \emph{Nucl.\ Instrum.\ Meth.\ A}, {\bf 798} (2015) 172-213, \arxiv{1502.01323}.
% \emph{The Pierre Auger Cosmic Ray Observatory},

\bibitem{bib:GDAS}
  Pierre Auger Collaboration, P.~Abreu \textit{et al.}, \emph{Astropart.\ Phys.}\ {\bf 35} (2012) 591--607, \arxiv{1201.2276}.
  %\emph{Description of Atmospheric Conditions at the Pierre Auger Observatory using the Global Data Assimilation System (GDAS)},
  
%\bibitem{bib:atmosphere}
%R.~Mussa for the Pierre Auger Collaboration, \emph{Nuc. Phys. B - Proc. Suppl.}, vol. 190, 272-277 (2009).

\bibitem{bib:fit_met} 
M.~Unger \textit{et al.}, \emph{Nucl.\ Instrum.\ Meth.\ A} {\bf 588} (2008) 433, \arxiv{0801.4309}.

\bibitem{bib:gh}
T.K.~Gaisser and A.M.~Hillas, \emph{Proc.\ 15th ICRC, Plovdiv, Bulgaria} {\bf 8} (1977) 353.

\bibitem{bib:flu_yield} M.~Ave \textit{et al.}, \emph{Astropart. Phys.} {\bf 42} (2013) 90.

\bibitem{bib:RL_first}
S.~Andringa, R.~Concei\c{c}\~{a}o and M.~Pimenta, \emph{Astropart.\ Phys.}\ {\bf 34} (2011) 360--367.

\bibitem{bib:j_matthew} 
J.A.J.~Matthews \textit{et al.}, \emph{J.\ Phys.\ G} {\bf 37} (2010) 025202, \arxiv{0909.4014}.

\bibitem{bib:hires2001}
The HiRes/MIA Collaboration, T.~Abu-Zayyad \textit{et al.}, \emph{Astropart.\ Phys.}\ {\bf 16} (2001) 1--11.

\bibitem{bib:hires}
G.~Hughes for the High Resolution Fly's Eye Collaboration, \emph{Proc.\ 30th ICRC, Merida, Mexico} {\bf 4} (2007) 405--408.

\bibitem{bib:proc_ruben}
R.~Concei\c{c}\~{a}o \textit{et al.}, \emph{24th European Cosmic Ray Symposium}, J. Phys. Conf. Ser. 632 (2015) no.1, 012087.

\bibitem{bib:xmax} 
Pierre Auger Collaboration, A.~Aab \textit{et~al.}, \emph{Phys.\ Rev.\ D} {\bf 90} (2014) 122005.

\bibitem{bib:atmos}
  Pierre Auger Collaboration, J.~Abraham \textit{et al.}, \emph{Astropart.\ Phys.}\ {\bf 33} (2010) 108--129, \arxiv{1002.0366}.
%\emph{A Study of the Effect of Molecular and Aerosol Conditions in the Atmosphere on Air Fluorescence Measurements at the Pierre Auger Observatory}, 
  
\bibitem{bib:clf}  
B.~Fick et. al., \emph{JINST}\ {\bf 1} (2006) P11003.
%\emph{ The Central Laser Facility at the Pierre Auger Observatory},

 \bibitem{bib:aero}
   Pierre Auger Collaboration, P.~Abreu \textit{et al.}, \emph{JINST}\ {\bf 8} (2013) P04009, \arxiv{1303.5576}.
%\emph{Techniques for Measuring Aerosols using the Central Laser Facility at the Pierre Auger Observatory}, 
   
\bibitem{bib:verziAtICRC13} 
V.~Verzi for the Pierre Auger Collaboration, \emph{Proc.\ 33rd ICRC, Rio de Janeiro, Brazil} (2013), \arxiv{1307.5059}.

\bibitem{bib:corsika} 
D. Heck \textit{et al.}, \emph{Report FZKA 6019}\ {\bf 26} (1998).

\bibitem{bib:eposlhc} T. Pierog and K. Werner,  \emph{Phys.\ Rev.\ Lett.}\ {\bf 101} (2008) 171101.

\bibitem{bib:qgsjetII04} S. Ostapchenko,  \emph{Phys.\ Rev.\ D}\ {\bf 83} (2011) 014018.

\bibitem{bib:sibyll23} F. Riehn \textit{et al.},  \emph{PoS(ICRC2017)}\ {\bf 301} (2017) \arxiv{1709.07227}.

\end{thebibliography}
\end{document}